# Spatial Non-Uniformity in Exfoliated $WS_2$ Single layers

I. Paradisanos[1,2], N. Pliatsikas[3], P. Patsalas[3], C. Fotakis[1,2], E. Kymakis[1,5],

G. Kioseoglou*[1,4], E. Stratakis*[1,4]

[1] Institute of Electronic Structure and Laser, Foundation for Research and Technology - Hellas, Heraklion, 71110, Crete, Greece

[2] Department of Physics, University of Crete, Heraklion, 71003, Crete, Greece

[3] Department of Physics, Aristotle University of Thessaloniki, Thessaloniki, 54124, Greece

[4] Department of Materials Science and Technology, University of Crete, Heraklion, 71003 Crete, Greece

[5] Center of Materials Technology and Photonics & Electrical Engineering Department, Technological Educational Institute (TEI) of Crete, Heraklion, 71004 Crete, Greece

## ABSTRACT

Monolayers of transition metal dichalcogenides (TMDs) are atomically thin two-dimensional crystals with attractive optoelectronic properties, which are promising for emerging applications in nanophotonics. Here, we report on the extraordinary spatial non-uniformity of the photoluminescence (PL) and strain properties of exfoliated $WS_2$ monolayers. Specifically, it is shown that the edges of such monolayers exhibit remarkably enhanced PL intensity compared to their respective central area. A comprehensive analysis of the recombination channels involved in the PL process demonstrates a spatial non-uniformity across the monolayer's surface and reflects on the non-uniformity of the intrinsic electron density across the monolayer. Auger electron imaging and spectroscopy studies complemented with PL measurements in different environments indicate that oxygen chemisorption and physisorption are the two fundamental mechanisms responsible for the observed non-uniformity. At the same time Raman spectroscopy analysis shows remarkable strain variations among the different locations of an individual monolayer, however such variations cannot be strictly correlated with the non-uniform PL emission. Our results shed light on the role of the chemical bonding on the competition between exciton complexes in monolayer WS$_2$, providing a way of engineering new nanophotonic functions using WS$_2$ monolayers. It is therefore envisaged that our findings could find diverse applications towards the development of TMDs-based optoelectronic devices.

**INTRODUCTION**

Single layers of transition metal dichalcogenides (TMDs) exhibit fundamentally new phenomena that can lead to novel 2D nanoelectronic and nanophotonic applications.[1-4] They are direct-gap semiconductors at the K-points of the Brillouin zone with a range of bandgaps in the visible regime. This makes them ideal candidates for a host of optoelectronic applications ranging from light-emitting diodes and field effect transistors to light harvesting, sensors, quantum state-metamaterials and electro-catalytic water splitting applications.[5-9] In addition, TMDs have shown properties such as nonblinking photon emission, high effective Young's modulus and spontaneous valley polarization.[10-12] Their optical properties are governed by strong excitonic transitions (neutral and charged) with very high binding energy for the neutral exciton of the order of 0.5-0.8 eV.[13-17] Based on their unique optical selection rules, these materials are also good candidates for the emerging field of valleytronics.[18-22]

The outstanding stretchability of TMDs is also promising for strain engineering and related applications. Indeed, the unprecedented energy-level tunability under strain could lead to devices with electronic properties that are controlled via mechanical deformation.[23] For example, single-layer MoS$_2$ is predicted to undergo a direct-to-indirect bandgap transition at ~ 2% of tensile uniaxial strain and a semiconducting-to-metallic transition at 10-15% of tensile biaxial strain.[24] Similar predictions are obtained for WS$_2$ monolayers.[25,26] Phonon softening and giant valley drift under strain have also been reported in MoS$_2$ and WS$_2$ monolayers.[27,28]

The boundaries of 2D TMDs crystals are reported to be non-atomically sharp and extremely susceptible to their environment, affecting not only the optical but also their transport properties.[29-31] Physisorbed molecules were shown to influence the PL properties, while annealing was drastically affecting the PL intensity.[32] These changes can be reversible as a result of charge transfer between the environment and the 2D crystal. Structural defects, mainly S-vacancies are also affecting the optical properties. Spatial non-uniformities in the PL emission energy and intensity have been previously observed but only in CVD grown single layers of $WS_2$. In those studies, they found a PL intensity increase, as well as a red shift in the emitted energy, as one moves from the central area to the edge of the sample. They attribute this to the larger local population of charge carriers at the edges and to the accumulation of excitons at those areas.[30,33,34]

Here we report on the extraordinary photoluminescence (PL) and strain properties, not only of the physical boundaries but also of the intentionally created ones via femtosecond laser ablation of mechanically exfoliated $WS_2$ monolayers.[35] In particular, there is a 3-fold enhancement of the PL intensity at the edges compared to the monolayer center, with the emission channels being of different origin. It should be emphasized that the term 'edge' used here is not associated with the monolayer's atomic edge, but with its peripheral areas. Contrary to previous reports, instead of a red shift in the PL energy, we see a systematic PL intensity transfer from the charged excitons that dominate the emission at the central area, to the neutral excitons at the edges of the exfoliated single layers. This clearly demonstrates a spatial non-uniformity of the intrinsic electron density across the monolayer surface. The relation of the observed properties with the chemical nature of the edges is analyzed via Scanning Auger Microscopy (SAM) and Auger Electron Spectroscopy (AES) with high spatial and analysis-depth resolution. The Auger analysis shows that Sulfur is replaced by Oxygen (at least in part) at the edges, an issue that was overlooked in the literature so far. Based on these findings, we demonstrate that chemisorbed Oxygen as well as physisorbed molecules (mainly $O_2$) affect the emission properties of $WS_2$ monolayers. Finally, it is shown that there are significant Raman shifts between different areas of the exfoliated $WS_2$ monolayers. The observed Raman shifts were attributed to inhomogeneous strain that varies among different surface areas and depends on the mechanical history of the sample during exfoliation.

**RESULTS AND DISCUSSION**

Micro-Raman spectroscopy with very low intensity in order to avoid structural damage was utilized to identify the number of layers in certain positions of the exfoliated flakes. The energy difference between the two most prominent Raman vibrational modes, $E^1_{2g}$ (in plane mode) and $A_{1g}$ (out of plane mode), is used extensively in the literature as the fingerprint of the layers' number. Specifically, 59-61 cm$^{-1}$ energy differences unambiguously confirm the existence of WS$_2$ monolayers.[36] Micro-photoluminescence (micro-PL) studies at 300K were performed using a setup in backscattering geometry, with a 543 nm continuous wave laser as an excitation source. Ablation experiments were performed using a fs Yb-doped potassium-gadolinium tungstate crystal laser system. With a microscope objective lens all laser beams were focused down to 1μm on the sample, placed on an XYZ translation stage, at normal incidence. SAM images and AES spectra were acquired using a system equipped with a field emission electron gun and a multichannel hemispherical sector electron analyzer in ultra-high vacuum conditions. More details on the sample preparation, optical characterization and Auger measurements are given at the Methods Section.

Figure 1a shows a typical optical microscopy image of exfoliated WS$_2$ flakes. The size of the monolayer region is 6–8μm across. When a defocused, spatially homogeneous 543 nm CW laser beam excites, in ambient conditions, the flake area, a robust photoluminescence (PL) is emitted solely from the monolayer (fluorescence image in Figure 1b). It is clear though, that the stronger PL emission is from the edges and extends laterally to about 1-1.5μm creating a "donut" effect. From the line scan in Figure 1b, an intensity profile was generated across the flake (Figure 1c) that reveals a 3-fold enhancement of the PL at the edges. This 3-fold enhancement is also evident in Figure 1d that compares the PL spectra at two different spots on the flake (inset of Figure 1d); spot A is from the center and spot B from the edge of the monolayer. Note here that points A and B, even though randomly chosen, are typical representative points from the central and peripheral areas of the monolayer. However, we observe similar PL spectra from different points of the same area (inner or edge). It can be observed that the PL emission from the central area is dominated by charged exciton (trion) recombination (X$^-$, at 1.97eV) while the emission from the edges is mainly due to the neutral excitons (X$^0$, at 2.01eV). The spectroscopic analysis provided in Figure 1d, sheds light on the origin of the recombination channels involved in the PL process. Charged versus neutral exciton emission demonstrates a spatial non-uniformity across the flake's surface and reflects on the non-uniformity of the intrinsic electron density across the monolayer. This electron density variation may stem from possible lattice changes due to physisorption and/or chemisorption processes of oxygen species that may take place and needs further investigation.

In order to identify the chemical variations along the bulk and monolayer $WS_2$ crystals, we employed SAM and AES. SAM is used to identify the extent of the $WS_2$ crystal and monolayer regions, as shown in Figure 2. In particular, we extracted SAM images from the $WS_2$ sample, which is shown in the optical microscopy image of Figure 2a. The corresponding SAM image, acquired using the kinetic energy of the $W_{MNN}$ Auger electrons, demonstrates the distribution of W over the surface as presented in Figure 2b. Apparently, the orange and red regions of Figure 2b correspond to bulk $WS_2$, while the green region is identified as the monolayer $WS_2$. The cyan and blue regions correspond to bare substrate; note that the signal-to-noise ratio of the monolayer $WS_2$ is indeed low, however, the SAM image is used only to identify the extent of the $WS_2$ monolayer and not to extract any chemical information. Based on the SAM image of Figure 2b we performed an AES line scan along the vertical dashed line with a spatial resolution of 50nm, in order to record the compositional and chemical variations along the line, with a better that signal-to-noise ratio than that of SAM, and in particular at the points A, B, C, that correspond to the bulk, the center, and the edge of the monolayer $WS_2$ respectively, and the point D that corresponds to the bare substrate.

In order to get more insight on the chemical bonding along the bulk and monolayer $WS_2$ we acquired detailed AES spectra of the $W_{MNN}$ and $S_{KLL}$. The $W_{MNN}$ (Figure 2c) peak is very sensitive to the chemical environment of W. Indeed, the $W_{MNN}$ peak is reported to be manifested at different kinetic energies for W-W bonds in metallic W (1729.6 eV), for W-S bonds in $WS_2$ (1728.5-1728.7 eV), and for W-O bonds in $WO_3$ (1723.8 eV).[37,38] The corresponding kinetic energy levels are presented in Figure 2c by vertical arrows. For the studied sample, pure W-S bonds in $WS_2$ were identified for the bulk $WS_2$ and for the center of the monolayer $WS_2$ (points A and B). On the contrary, for the edge of the monolayer $WS_2$ (point C) there is a clear shift to lower kinetic energy, which indicates the emergence of the W-O bonds. Given that the corresponding $S_{KLL}$ peaks (Figure 2d) do not show any spectral variation, and consequently no variation of the bonding of S, we may safely conclude that S is replaced by O (oxidizing in part the W and forming locally $WO_2$, in which W is in +4 oxidation state in contrast to +6 of $WO_3$, an assessment based on the intermediate value of the AES peak reported for point C) at the edges of the monolayer $WS_2$. In order to verify that this is not an accidental observation, we repeated independently the SAM and AES analysis for a second $WS_2$ sample (see the on-line supplemental information, Figures S4,S5).

The result from the Auger spectroscopy that oxygen replaces sulfur at the edges is really significant and it is a direct experimental verification of the theoretically predicted chemisorption of oxygen that takes place on the $WS_2$ monolayer surface.[39,40] In the study by Liu et al., a chemical bond between oxygen and the metal atoms of TMDs was predicted indicating that oxygen was chemically adsorbed onto the defective sites of the layers. During this process, oxygen bonds to the unsaturated Mo or W and gains electrons from the metal atom leading to an electron depletion from the layer. Liu et al. found that chemisorbed $O_2$ has a strong impact that can change the monolayer sulphur vacancies from harmful carrier-traps to electronically benign sites.[40] Their calculation shows that electrons accumulate at chemisorbed $O_2$ by depleting those previously present at vacancy sites. A barrier of 0.97eV must be overcome before reaching the chemisorbed state from the physisorbed state, which can easily be overcome with our 543nm (2.28eV) excitation laser. Based on these findings we postulate that the enhanced oxygen chemisorption at the monolayer peripheral regions can be possibly attributed to oxygen diffusion from the atomic edges to inner areas or to an increase of sulphur vacancies in such areas due to mechanical exfoliation, which is inherently a harsh process. However, further studies are required to clarify this issue. Therefore the chemisorbed oxygen, acting as an electron receptor, changes the local intrinsic charge density and thus affecting the recombination channels. The fluoresence "donut" effect of Figure 1b as well as the differences in the recombination channels observed in Figure 1d (charged vs neutral exciton emission) are direct consequences of this spatial non-uniformity of the intrinsic electron density across the monolayer surface.

Chemisorption may not be the only process that take place on the air-exposed monolayer surfaces. Indeed, physisorption has been already shown to play a significant role on the emission channels of $WS_2$ single layers.[31] A DFT calculation has shown that oxygen can be psysisorbed on the sulfur vacancy sites.[32] To check the effect of the psysisorbed species on the PL, we study the emission from different spots (inner and edge) from another flake (Figure 3a) in three different enviroments; air, vacuum and nitrogen. In Figure 3b we compare the normalized PL spectra at four different locations of $WS_2$ monolayer, consisting of two inner (numbers 1 and 3) and two edge (numbers 2 and 4) points in ambient conditions. There is a clear neutral exciton emission from the edges with respect to inner regions where the emission from charged excitons is prominent. By placing now the samples in low- vacuum conditions (Figure 3c), the PL spectra are dominated by the charged excitons (around 1.96-1.97 eV) regardless the position on the sample. However, comparing the PL spectra at points 2 (edge, blue curve) and 3 (center, pink curve) it is clear that there is an enhancement of the neutral exciton (2.01 eV) emission intensity even though the charged exciton is still the dominant one. We also performed measurements in 600 mbar nitrogen conditions, as displayed in Figure 3d. In this case, we see a similar behavior as the PL in vacuum. In other words, even though the main emission channel originates from the recombination of trions, there is an appreciable enhancement of the neutral exciton intensity at the edges compared to the center of the flake. Supposing that chemisorption was the only process taking place, PL emission in vacuum and nitrogen should have been the same as in air. This is not the case as Figure 3 clearly shows. Therefore, our results demonstrate that apart from oxygen chemisortpion at the edges (shown by Auger measurements), oxygen physisorption processes also play a significant role in the emission properties.

To further confirm the enhanced oxygen concentration at the monolayer edges compared to its center, we monitored the integrated PL intensity from the edges and inner locations as a function of the pre-irradiation time of the sample in air, as shown in Figure 4. The x-axis represents the time that a specific location is exposed to laser irradiation before the acquisition of the PL spectrum. In all measurements, the collection time of the PL signal was kept constant. It is observed that while the PL signal from the edges exhibits a monotonic decrease up to 65 seconds of pre-irradiation time, the corresponding PL intensity from the inner regions quickly drops to a saturated value within only 15 seconds of pre-irradiation time. The observed delay for the PL intensity stabilization at the edges, compared to the inner areas, is another indication for the presence of higher concentration of oxygen species at the edges. Notice also that after 65 seconds of pre-irradiation time when most of the psysisorbed oxygen has been removed from the monolayer surface, the PL intensity from the edge region is still higher than the corresponding one from the inner area. This result strongly supports the findings of Auger spectroscopy presented in Figure 2c. Therefore, both oxygen chemisorption at the edges, as well as oxygen physisorption across the whole surface of the monolayer have an effect on the intrinsic electron density and subsequently the emission recombination channels.

Strain could also have an effect in the optical properties of semiconductors since it affects the band structure. To examine such a possibility for the observed non-uniformity in the PL (inner vs edge areas), Raman spectroscopy at different positions within a monolayer area has been performed. Figure 5a shows an optical microscopy image of a $WS_2$ monolayer, where different positions (from 1 to 13) used to obtain Raman spectra are indicated. Specifically, numbers 1 to 4 represent the upper edge region of the single layer, numbers 5 to 8 are located in the inner area and numbers 9 to 13 correspond to the lower monolayer edges. A typical Raman spectrum is shown in Figure 5b. The energy difference between the two most prominent Raman vibrational modes, $E^1_{2g}$ (in plane) and $A_{1g}$ (out of plane), is used extensively in the literature as the fingerprint of the number of layers. In all cases, this difference was ranging from 58.8 cm$^{-1}$ to 61.3 cm$^{-1}$, (Figure S1, supplementary information) which confirms the presence of a monolayer in all locations studied. As shown in Figures 5c and 5d, the biggest energy shift in both the in- and out-of plane modes, was observed for points 1 and 13 (the two edge positions on the left side). Figure 5e presents the Raman shift values of $A_{1g}$ and $E^1_{2g}$ modes for the 13 different positions. For the analysis of the $E^1_{2g}$ mode position we have considered the contribution of the second order 2LA mode by fitting the experimental data using a Voigt spectral profile (Fig. S6). There is a noticeable systematic red or blue shift in both vibrational modes at the monolayer edges (positions 1-4 and 9-13), compared to its inner part (positions 5 to 8). Differences ranging from 0.5 cm$^{-1}$ (resolution limit) to a maximum of about 4 cm$^{-1}$ for the $A_{1g}$ mode and 3.2 cm$^{-1}$ for the $E^1_{2g}$ mode between points 1 and 13, can be observed. These notable shifts of the two main vibrational modes can also be seen in Figure S2 (supplementary information), where the energy of $A_{1g}$ is plotted as a function of the energy of $E^1_{2g}$, for all the different points measured. It should be noted that higher concentration of oxygen species at the edges could lead to variations of the Raman modes due to the difference in mass between oxygen and sulfur. However, in that case such variations would normally exhibit the same type of shift at both edges of the sample which is not the case according to Figure 5e; Raman modes at points 1-4 are red-shifted with respect to the center while the modes of points 9-13 are blue-shifted. Additional experiments on a different sample showing opposite Raman shifts between different edges of the same monolayer are presented in Figure S3 (supplementary information).

Based on the way the monolayers are prepared, i.e. by mechanical exfoliation, it is highly possible that they experience irregular mechanical forces that result to inhomogeneous strain across the monolayer's area. Thus, inhomogeneous strain can explain the observed behaviour of the $E^1_{2g}$ and $A_{1g}$ modes. We cannot exclude though the synergy of doping effects on the variations of the $A_{1g}$ mode, considering the well-reported sensitivity of this mode on charging effects.[41] However, it is very difficult to quantitatively assess the respective contributions, considering that it is hard to decouple these two synergetic effects. Hence, based on our available data, it can be postulated that there is no one-to-one correlation of the observed PL non-uniformity with the Raman signal variations and oxygen chemisorption and physisorption remain the two prominent mechanisms responsible for the spatial non-uniformity of the electron density across the monolayer surface.

It should be finally mentioned that the observed PL enhancement at the $WS_2$ monolayer edges was also observed after the intentional creation of new peripheral edges, via femtosecond laser ablation of a part of the monolayer in ambient conditions. This is depicted in Figure 6a, for the sample shown in Figure 1b, while Figure 6b shows a representative line scan of the respective fluorescence image through the ablated area. The above observation further confirms the important role of the edges on the emission properties of $WS_2$ monolayer. It also provides a simple approach to form arbitrary patterns of enhanced PL emission onto the monolayer surface.

**MATERIALS AND METHODS**

All samples used in this study were mechanically exfoliated from bulk crystals (HQ Graphene) and deposited onto Si/Silicon Oxide (290 nm) wafers. The typical size of monolayer regions are 6–8 μm across and were identified with an optical microscope and confirmed with Raman spectroscopy at room temperature and 473nm excitation. Raman spectroscopy (Thermo Scientific) has been established as a reliable tool for determining the specific number of layers in transition metal dichalcogenides.[36,42] In $WS_2$, the 60-61 $cm^{-1}$ energy difference between the two main vibrational modes, $E^1_{2g}$ at 357 $cm^{-1}$ and $A_{1g}$ at 417 $cm^{-1}$, is the fingerprint for the accurate determination of the single layer.

Micro-photoluminescence studies at 300K were performed in a backscattering geometry, with a 543 nm continuous wave laser as an excitation source. Ablation experiments were performed using a 175 femtosecond (fs) Yb-doped Potassium-Gadolinium Tungstate crystal laser system, operating at 513 nm wavelength. The energy of the beam was controlled via a combination of a waveplate, a linear polarizer and a series of neutral density filters. An iris aperture was used to obtain the central part of the beam and acquire a uniform energy distribution. Using a microscope objective lens, all laser beams were focused down to 1μm on the sample, placed on an XYZ translation stage, at normal incidence. In a typical experiment the fs laser beam ablated a part of monolayers, via focusing of the beam onto their edges. The alignment and irradiation processes could be continuously monitored by means of a CCD imaging setup.

Scanning Auger Microscopy (SAM) images and Auger Electron Spectra (AES) were acquired in a Kratos Axis Ultra DLD system equipped with a field emission electron gun, a hemispherical sector analyzer and a multichannel electron detector in ultra-high vacuum conditions (Pb=<$2\times10^{-10}$ mbar). No sputter etching was used to clean the surface, in order to prevent any surface damage, as well as any alteration of the chemistry of the monolayer $WS_2$. Despite the lack of surface cleaning, no oxygen ($O_{KLL}$ peak) was detected on the surface of the $WS_2$ crystal, proving the chemical inertness of its surface, when the sample is in multilayer crystal form.

**CONCLUSIONS**

In conclusion, the spatial non-uniformity of the PL emission and lattice strain properties of $WS_2$ monolayers, exfoliated from the natural crystal, has been demonstrated. We showed that the PL emission is significantly enhanced at the monolayer's edges compared to their inner parts. We attributed such PL enhancement to the pronounced oxygen chemisorption and physisorption at the edges, which affects the spatial distribution between exciton complexes in monolayer $WS_2$, giving rise to spatial non-uniformity of the electron density across the monolayer surface. In contrast, we showed the emission non-uniformity could not be clearly correlated with the lattice strain variations. We envisage that our observations of the spatial non-uniformity of the emission and lattice strain properties will be useful towards engineering of specialized electronic devices based on $WS_2$ and other TMDs monolayers. In addition, a simple approach to form arbitrary patterns of enhanced PL emission onto the monolayer surface via femtosecond laser ablation is proposed.

**ACKNOWLEDGEMENTS**

This work is supported by the European Research Infrastructure NFFA-Europe, funded by EU's H2020 framework programme for research and innovation under grant agreement n. 654360.

**FIGURES AND FIGURE CAPTIONS**

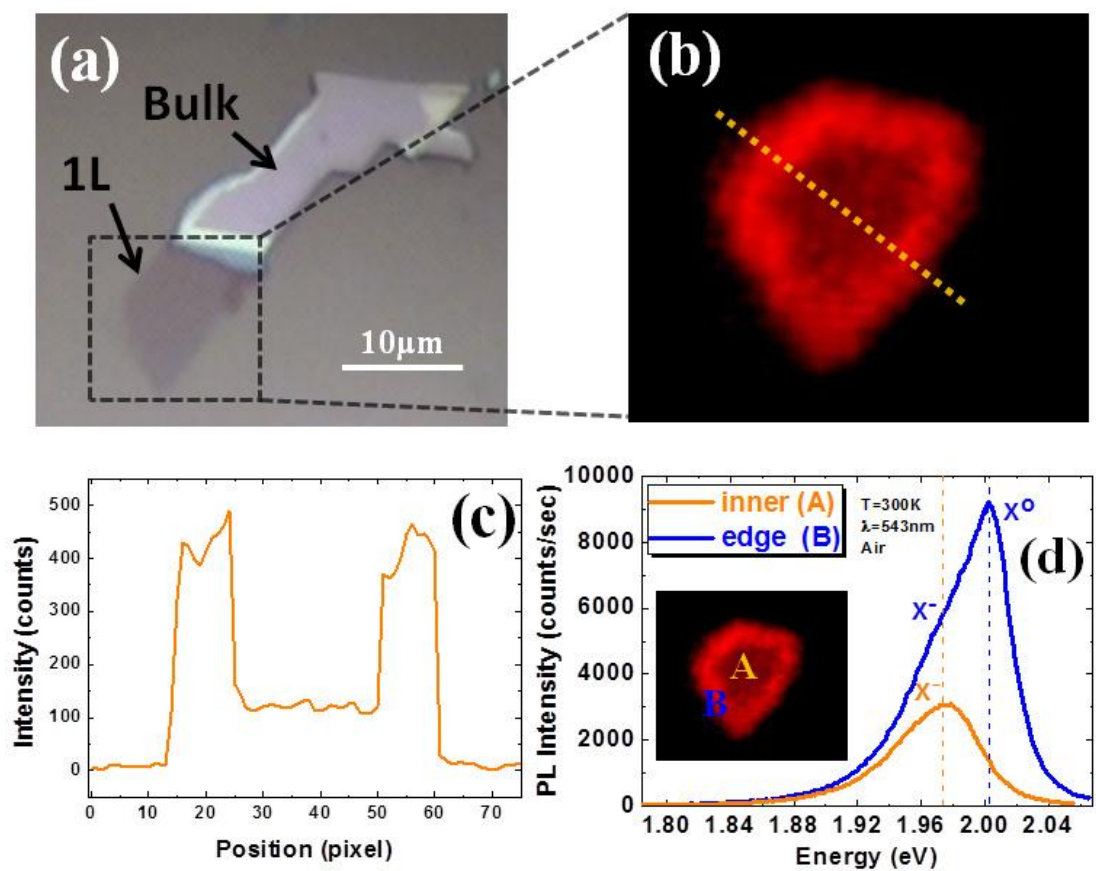

**Fig.1** (a) Typical optical microscopy image of exfoliated $WS_2$ flakes. The size of the monolayer region is 6–8 μm across. (b) Fluorescence image of $WS_2$ monolayer using a spatially homogeneous 543nm He-Ne laser beam. The stronger PL emission is from the edges, creating a "donut" effect. (c) Intensity profile across the flake, following the line scan of Fig. 1b. A 3-fold enhancement of the PL at the edges is revealed. (d) Representative PL spectra comparison of two different spots on the flake (inset of Figure 1d); spot A is from the center and spot B from the edge of the monolayer. The PL emission from the central area is dominated by charged exciton (trion) recombination ($X^-$, at 1.97eV) while the emission from the edges is mainly due to the neutral excitons ($X^0$, at 2.01eV).

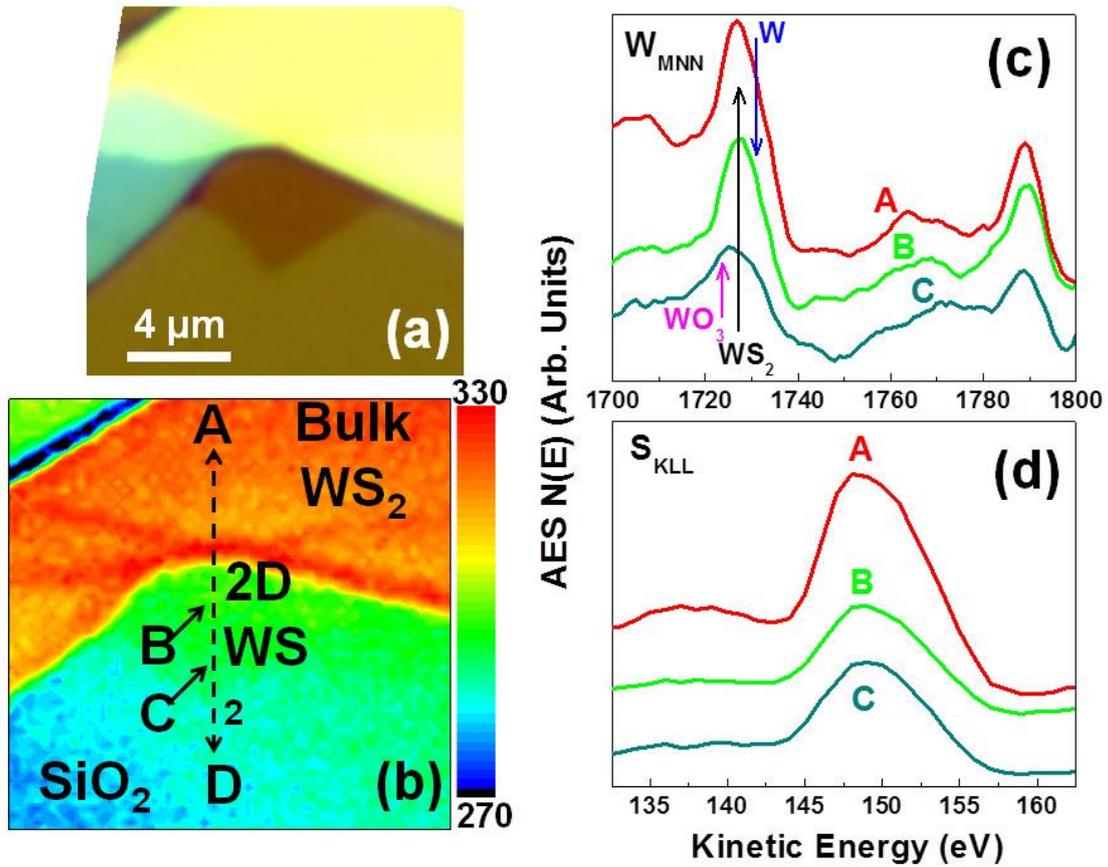

**Fig.2** (a) Optical microscope image of a $WS_2$ crystal (yellow area) with a $WS_2$ monolayer extended beyond the crystal (dark brown area), (b) the surface distribution of W in the same region acquired by SAM recording the $W_{MNN}$ peak strength; the vertical dashed line indicates the line-scan path, while A,B,C and D are the points where AES spectra have been measured (and correspond to bulk $WS_2$, the center of monolayer $WS_2$, the edge of the monolayer $WS_2$ and bare $SiO_2$, respectively), (c) $W_{MNN}$ AES spectra from points A,B and C; the vertical arrows indicate the exact literature values for W-W, W-S and W-O (in $WO_3$) bonds, and (d) $S_{KLL}$ AES spectra from points A, B and C.

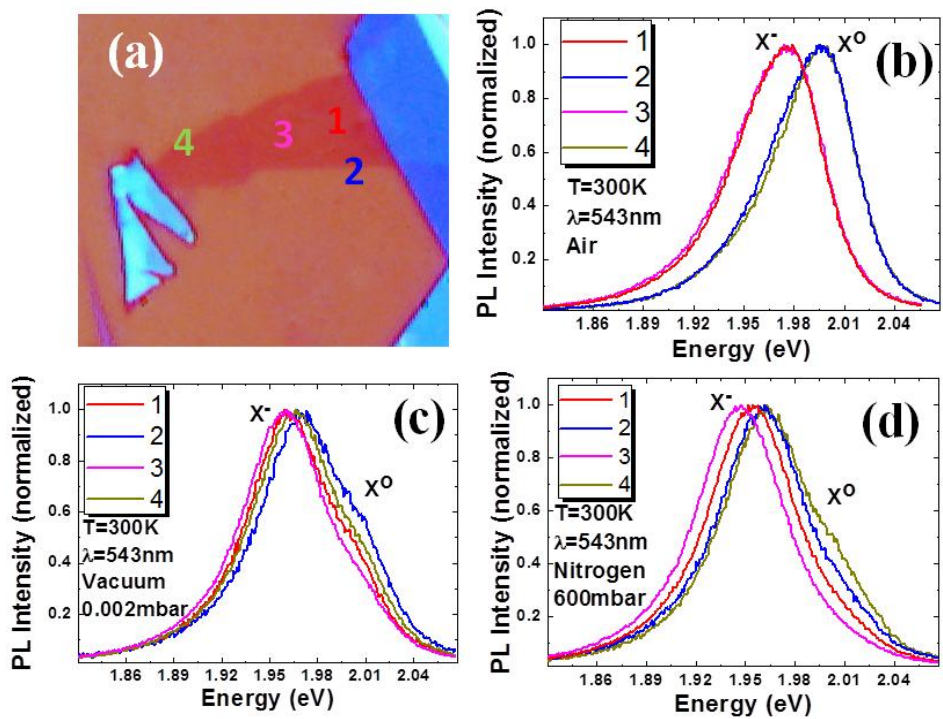

**Fig.3** (a) Optical microscopy image of WS$_2$ monolayer indicating two inner (numbers 1 and 3) and two edge (numbers 2 and 4) points. (b) Normalized PL spectra of four different locations of WS$_2$ monolayer, consisting of two inner (numbers 1 and 3) and two edge (numbers 2 and 4) points in ambient conditions. There is a clear neutral exciton emission from the edges with respect to inner regions where the emission from charged excitons is prominent. (c) Normalized PL spectra of the same four different locations in low-vacuum conditions. By comparing the PL spectra at points 2 (edge, blue curve) and 3 (center, pink curve) it is clear that there is an enhancement of the neutral exciton (2.01 eV) emission intensity even though the charged exciton is still the dominant one. (d) Normalized PL spectra in 600 mbar nitrogen conditions. A similar behavior as the PL in vacuum is observed.

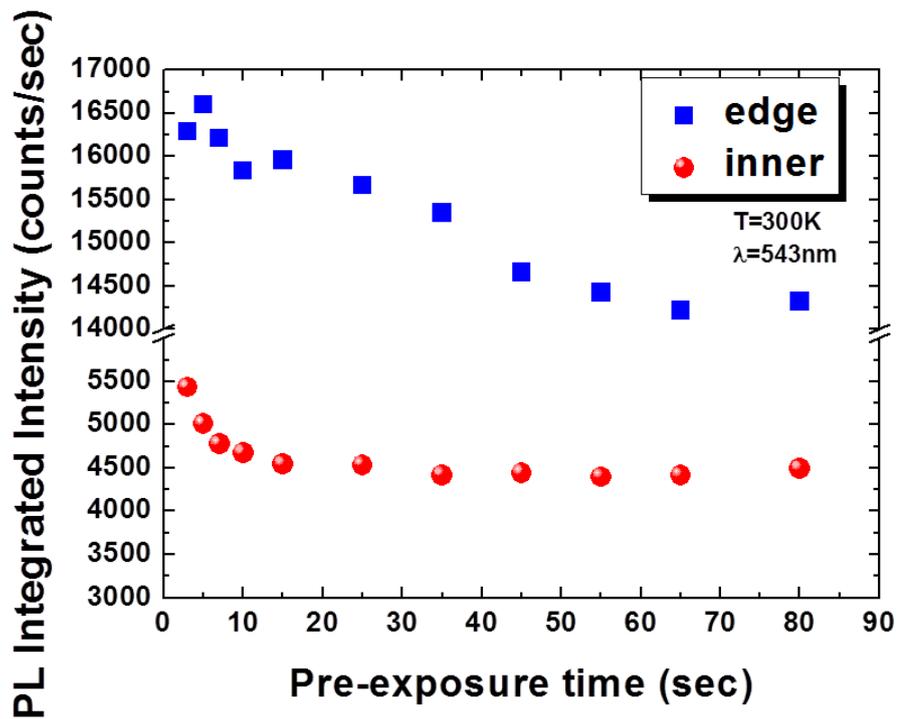

**Fig.4** Integrated PL intensity from the edges and inner locations as a function of the pre-irradiation time of the sample in air. The x-axis represents the time that a specific location is exposed to laser irradiation before the acquisition of the PL spectrum. In all measurements, the collection time of the PL signal was kept constant. While the PL signal from the edges exhibits a monotonic decrease up to 65 seconds of pre-irradiation time, the corresponding PL intensity from the inner regions quickly drops to a saturated value within only 15 seconds of pre-irradiation time.

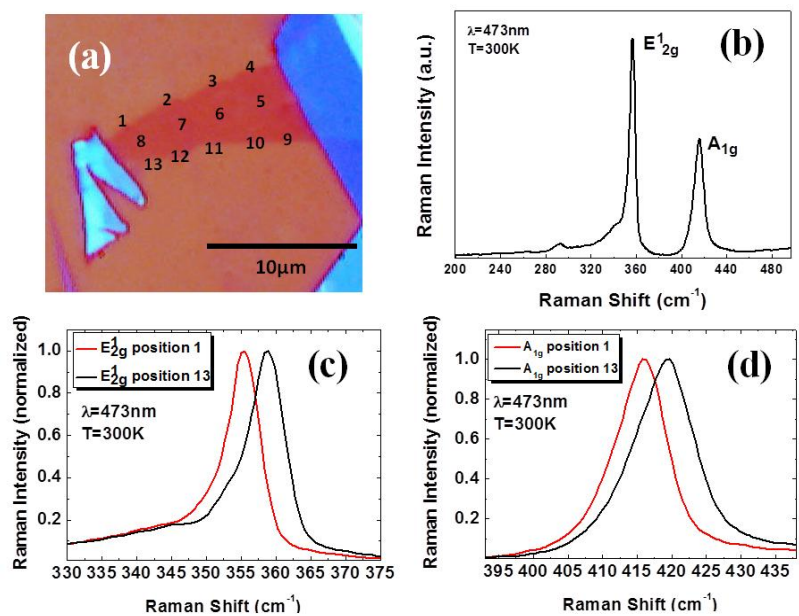

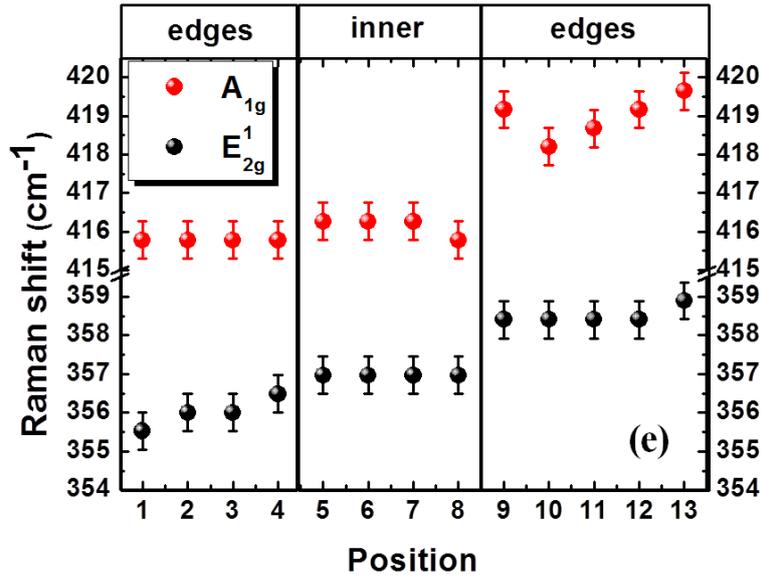

**Fig.5** (a) Optical microscopy image of a $WS_2$ monolayer, where different positions (from 1 to 13) used to obtain Raman spectra are indicated. Numbers 1 to 4 represent the upper edge region of the single layer, numbers 5 to 8 are located in the inner area and numbers 9 to 13 correspond to the lower edges of the monolayer. (b) Typical Raman spectrum of $WS_2$ monolayer. The energy difference between the two most prominent Raman vibrational modes, $E^1_{2g}$ (in plane) and $A_{1g}$ (out of plane), is used extensively in the literature as the fingerprint of the number of layers. In all cases, this difference was ranging from 58.8 cm$^{-1}$ to 60.8 cm$^{-1}$, which confirms the presence of a monolayer in all locations studied. The biggest energy shift with respect to the values at the center, in both the in- (c) and out-(d) of plane modes, was observed for points 1 and 13 (the two edge positions on the left side). (e) Raman shift values of $A_{1g}$ and $E^1_{2g}$ modes for the 13 different positions. There is a noticeable systematic red or blue shift in both vibrational modes at the monolayer edges (positions 1-4 and 9-13), compared to its inner part (positions 5 to 8). Differences ranging from 0.5 cm$^{-1}$ to a maximum of 4 cm$^{-1}$ for the $A_{1g}$ mode and 3.5 cm$^{-1}$ for the $E^1_{2g}$ mode, can be observed.

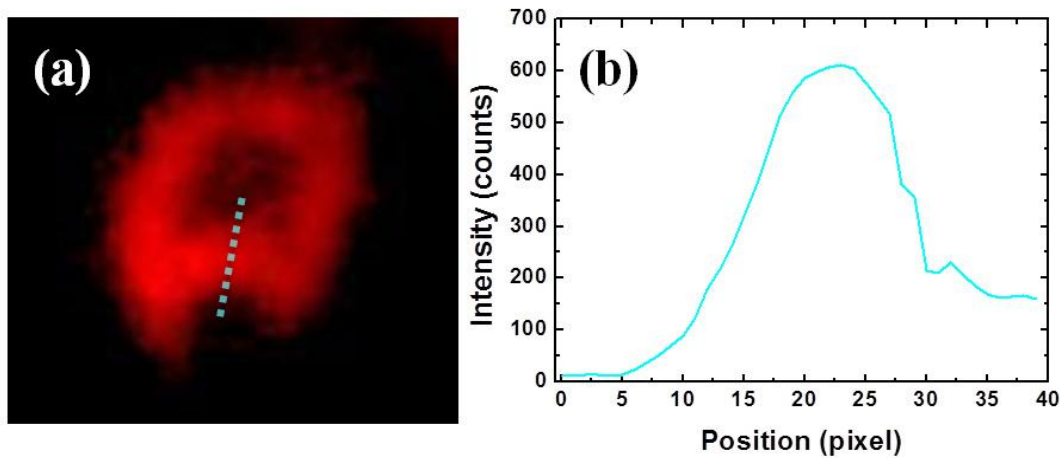

**Fig.6** (a) Optical microscopy image showing the ablated part of the monolayer (sample of Figure 1b) in ambient conditions. PL enhancement at the $WS_2$ monolayer edges is observed even after the intentional creation of new peripheral edges via femtosecond laser ablation. (b) Representative line scan of the respective fluorescence image through the ablated area. This observation provides a simple approach to form arbitrary patterns of enhanced PL emission onto the monolayer surface.

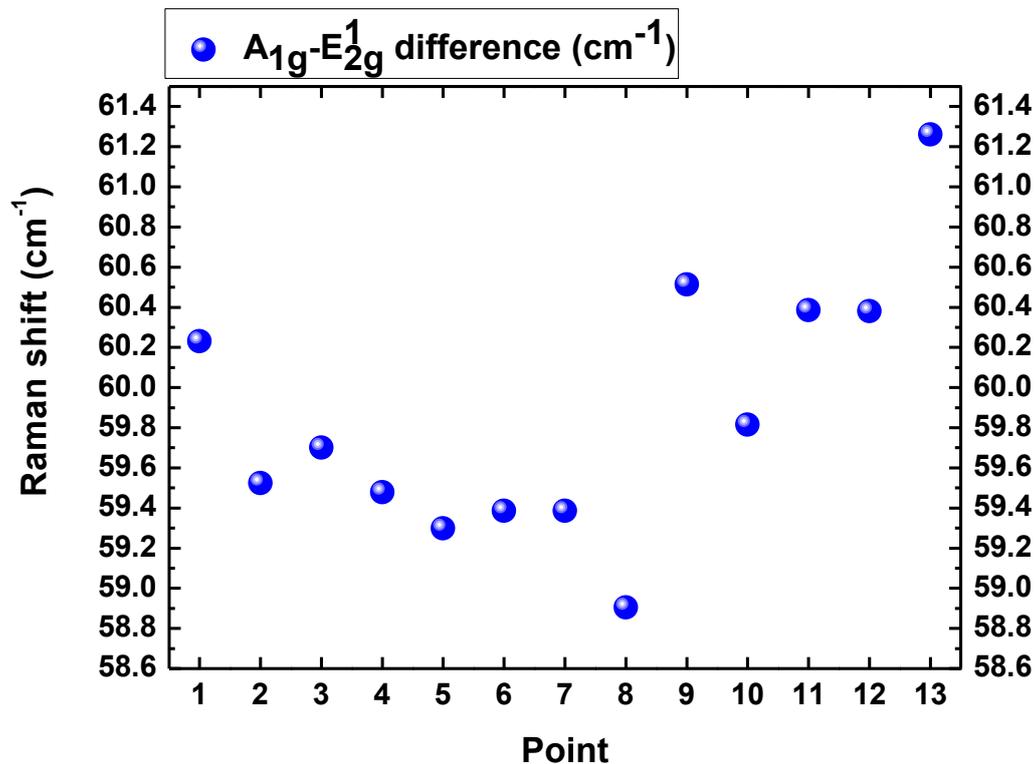

**Fig. S1** Energy difference between the two most prominent Raman vibrational modes, $E_{2g}^1$ (in plane) and $A_{1g}$ (out of plane), for the 13 different points of the $WS_2$ monolayer. This difference is ranging from 58.8 cm$^{-1}$ to 61.3 cm$^{-1}$.

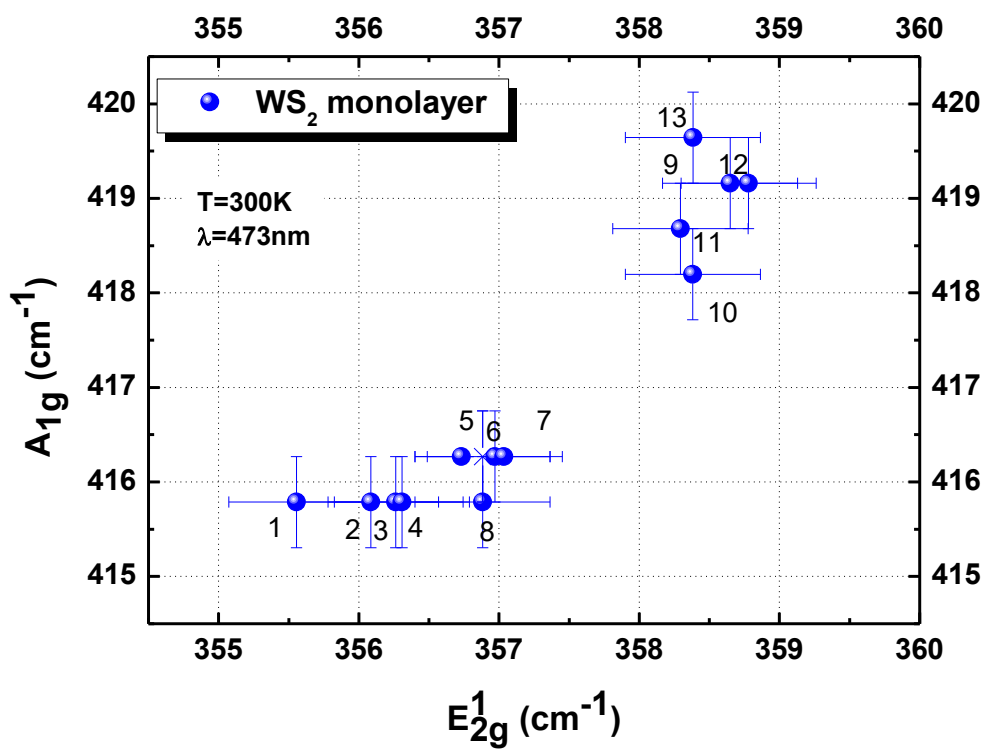

**Fig. S2** Energy of $A_{1g}$ as a function of the energy of $E^1_{2g}$, for all the different points measured. Differences ranging from 0.5 cm$^{-1}$ (resolution limit) to a maximum of 4 cm$^{-1}$ for the $A_{1g}$ mode and 3.2 cm$^{-1}$ for the $E^1_{2g}$ mode, can be observed. Two different groups of points can be distinguished (1 to 8 and 9 to 13) suggesting strain heterogeneity across the monolayer's area.

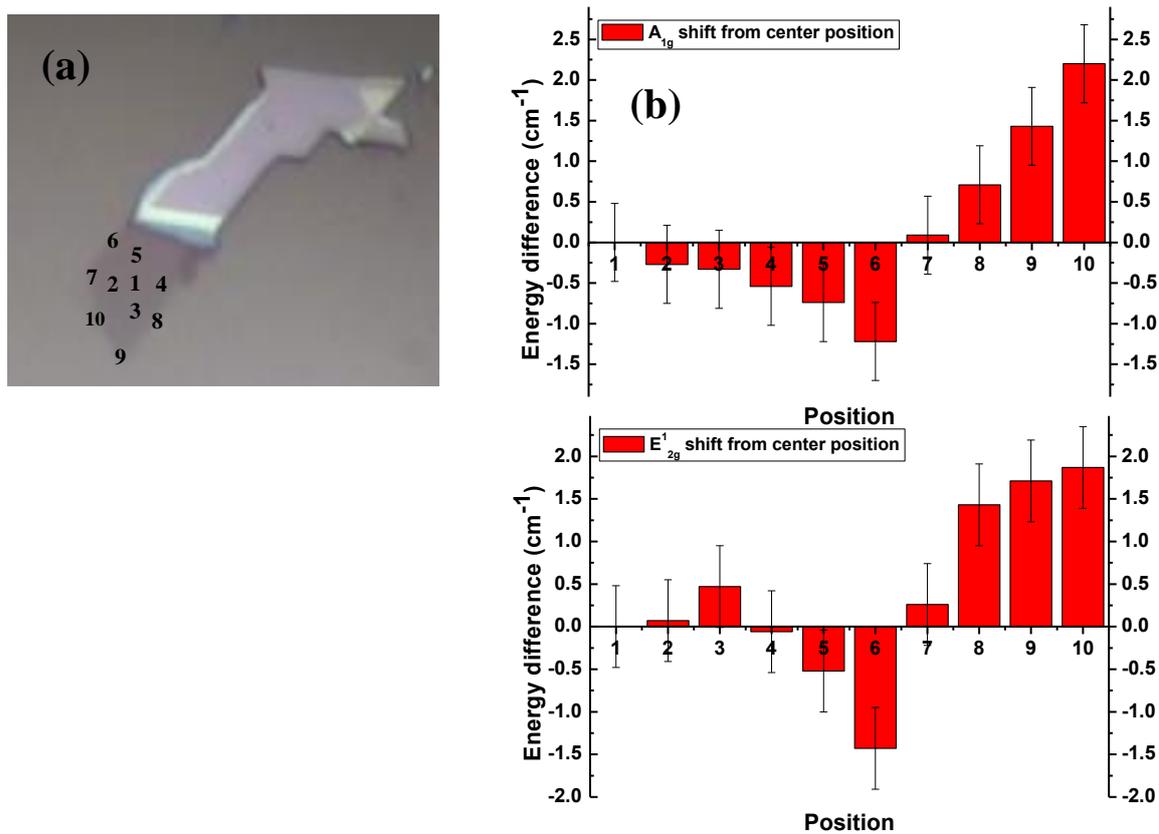

**Fig. S3** (a) Optical microscopy image of a WS2 monolayer, where different positions (from 1 to 10) used to obtain Raman spectra are indicated. Specifically, number 6 to 10 represent the edge regions of the single layer and numbers 1 to 5 are located in the inner area. (b) Raman shifts of the different areas with respect to position 1 that is located in the center of the monolayer.

In order to confirm that the observations of **Fig. 2** regarding the partial oxidation of the edge of the ML $WS_2$ we repeated the SAM/AES analysis to second sample, as shown in **Fig. S4**. SEM and SAM revealed an area between the bulk and ML $WS_2$, where the sample is completely detached (dark area in SEM, cyan area in SAM). After identifying the extent of ML $WS_2$ by SEM we acquired AES spectra at points A, B and C that correspond to bulk $WS_2$, the center of monolayer $WS_2$ and the edge of the monolayer $WS_2$, respectively. The shift of the principal $W_{MNN}$ peak to lower kinetic energy, indicating the oxidation of W, is also clearly observed for this second sample.

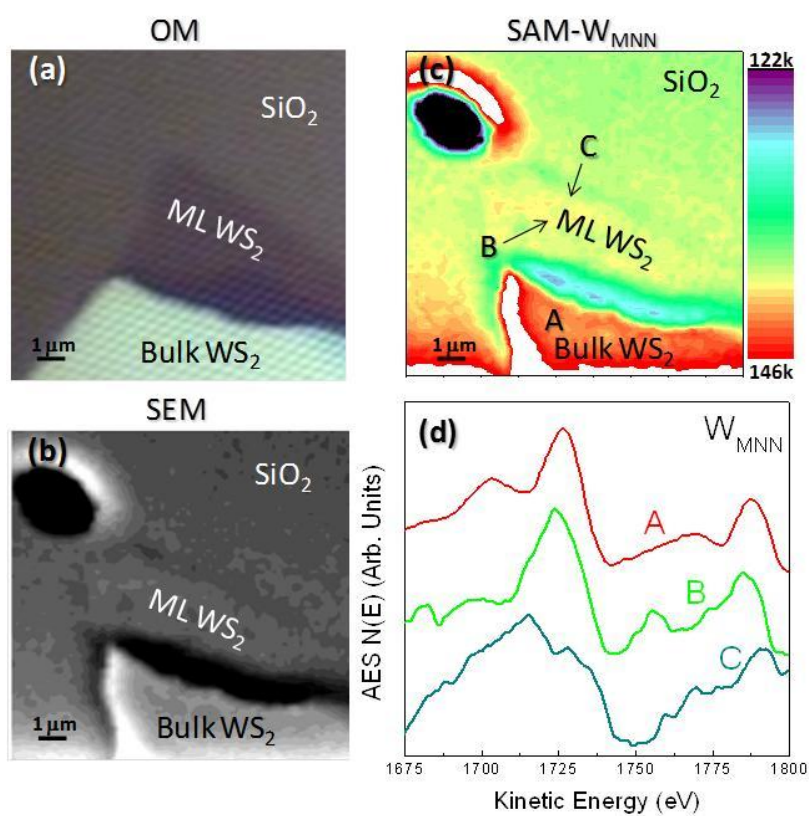

**Fig. S4** (a) Optical microscope image of a $WS_2$ crystal (bright area) with a $WS_2$ monolayer extended beyond the crystal (dark shadowed area), (b) SEM image from the same area where the ML $WS_2$ is better resolved, (c) the surface distribution of W in the same region acquired by SAM recording the $W_{MNN}$ peak strength; A, B, and C are the points where AES spectra have been measured (and correspond to bulk $WS_2$, the center of monolayer $WS_2$ and the edge of the monolayer $WS_2$), (d) $W_{MNN}$ AES spectra from points A, B and C.

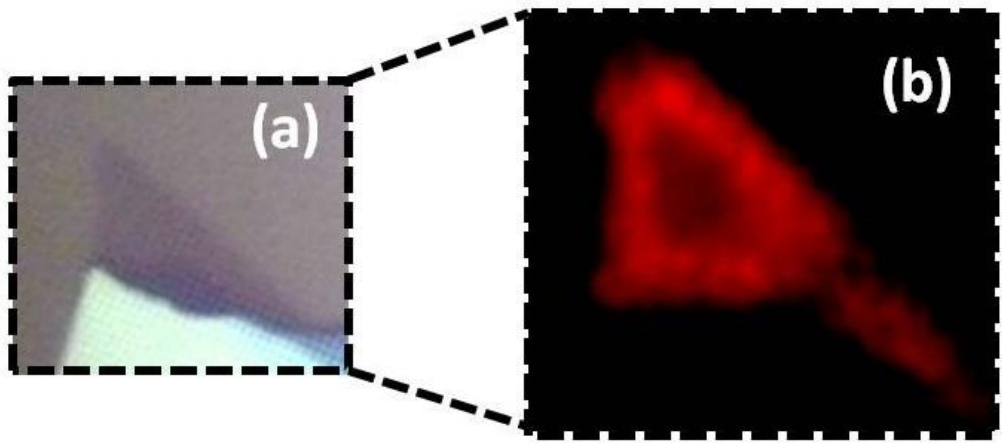

**Fig. S5** (a) Typical optical microscopy image of exfoliated $WS_2$ monolayer of Fig. S4. (b) Fluorescence image of $WS_2$ monolayer using a spatially homogeneous 543nm He-Ne laser beam.

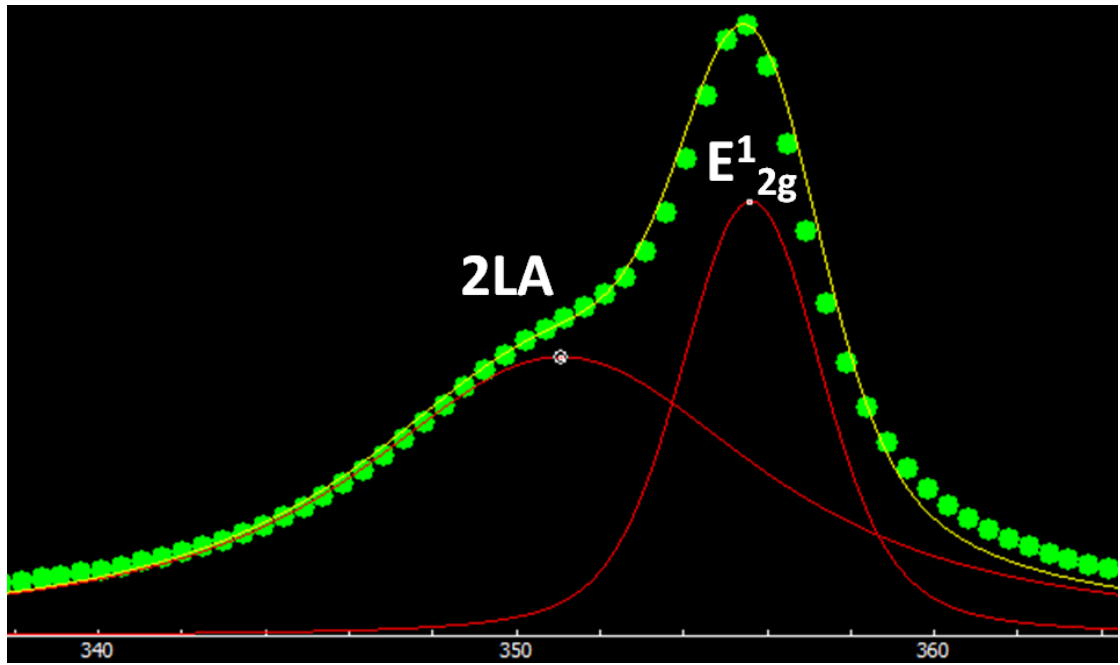

**Fig. S6** Fitting result of $E^1_{2g}$ and 2LA modes. The green spots correspond to the experimental data, red curves represent the voigt function of each mode and yellow line is the sum of the two functions.